\documentstyle{article}\begin{document}
\title{Relativistic diffusion of  elementary particles with spin}
\author{ Z. Haba\\
Institute of Theoretical Physics, University of Wroclaw,\\ 50-204
Wroclaw, Plac Maxa Borna 9, Poland\\
email:zhab@ift.uni.wroc.pl}\date{\today}\maketitle
\begin{abstract}
We obtain a  generalization of the relativistic diffusion of Schay
and Dudley for particles with spin. The diffusion equation is
a classical version of an equation for the Wigner
function of an elementary particle. The elementary particle is
described by a unitary irreducible representation of the Poincare
group realized in the Hilbert space of wave functions in the
momentum space. The arbitrariness of the Wigner rotation appears
as a gauge freedom of the diffusion equation. The spin is
described  by an $SU(2)$ connection of a fiber bundle over the momentum hyperbolic
space (the mass-shell). Motion in an electromagnetic field,
transport equations and equilibrium states are discussed.
\end{abstract}
 \section{Introduction}
 Approximations leading  to a diffusive motion seem to apply
 to non-relativistic as well as relativistic particles.
 Nevertheless, the problem of a relativistic extension of the diffusion theory
 encounters
some difficulties. There have been various approaches to a
solution of the problem ( see \cite{lop}
\cite{hakim}\cite{deb}\cite{dunkel}\cite{schay}\cite{dudley}\cite{hor1};
for a review and further references see
\cite{talkner}\cite{deb2}). In our earlier papers
\cite{haba}-\cite{haba2} we followed the approach of Schay
\cite{schay} and Dudley \cite{dudley} who defined a relatvistic
analog of the Brownian motion on the phase space as the diffusion
preserving the mass-shell in the momentum space. We think that one
should approach the relativistic dynamics starting from quantum
field theory which describes multiparticle interactions in a
relativistic way. In \cite{haba}\cite{haba2} we have discussed the
approach to equilibrium, the friction terms, the transport
equations  and an interaction with electromagnetic field. In
another approach, based on a relativistic Wigner function
\cite{zachar}\cite{elze}, transport equations for relativistic
particles have been derived from relativistic quantum field
theory. These transport equations should have a diffusion equation
as a classical non-relativistic limit.

We are interested in a classical diffusion which would approximate
relativistic motion of a quantum particle with spin. The Wigner
function is the quantum analog of the phase space distribution.
The (quantum) Wigner function in general is not positively
definite on the phase space although it can be positively definite
on the position space or momentum space. We have shown in
\cite{haba2} that Schay and Dudley definition of the diffusion
results  from a quantum Wigner function which is positive.
  We have derived the relativistic diffusion
of spinless particles \cite{haba2} from a quantum master equation
describing an evolution in the proper time (a relativistic master
equation in  proper time is discussed also in \cite{hor2})

In this paper we define a dissipative Lorentz invariant dynamics
of elementary particles described by a positively defined
probability distribution. This is an extension of the diffusion of
spinless particles to particles with spin. We look for a second
order Lorentz invariant differential operator on the mass-shell as
a generator of the diffusion. Our starting point is an observation
that the diffusion of Schay \cite{schay} and Dudley\cite{dudley}
as well as the diffusion of spinless massless particles
\cite{haba2} is generated by the operator $M_{\mu\nu}M^{\mu\nu}$
constructed from the spin zero representation of generators
$M_{\mu\nu}$ of the Lorentz group. An elementary particle is
described by a unitary irreducible representation of the Poincare
group. In the Hilbert space of wave functions defined on the
momentum space the representation is realized by an irreducible
representation of the little group of Wigner rotations. There is
an arbitrariness in the definition of the Wigner rotation
resulting from the non-uniquness of the boost
($(m,0,0,0)\rightarrow p$). As a consequence the generators of the
Lorentz rotations depend on an arbitrary function $g(p)\in SU(2)$.
The resulting diffusion equation also depends on the choice of the
boost transformation. We show that the transformation of the
diffusion equations can be interpreted as a gauge transformation.
A gauge invariant description of the diffusion of particles with
spin is possible if spin is identified with a  curvature of an
$SU(2)$ vector bundle over the hyperbolic space (the mass-shell).
The gauge invariance is used as a guiding rule for a derivation of
the relativistic diffusion equation in an electromagnetic field.

In an equivalent description of spin we consider a phase space
which is a product of the particle phase space with a sphere. The
diffusion of a particle with spin can be described as a diffusion
on the extended phase space (without any discrete variables).
Then, an evolution of discrete projections $\sigma $ of spin is
defined by a transition function
$P_{\tau}(x,p,\sigma;x^{\prime},p^{\prime},\sigma^{\prime})$ which
is a matrix element of the transition function on the extended
phase space .

We discuss transport equations resulting from an elimination of
the proper time and equilibrium distributions reached at infinite
laboratory time. Applications of the spin diffusion to a
description of relativistic entanglement, decoherence and
magnetization are briefly discussed.
\section{Wigner wave
functions of elementary particles } Let $A\in SL(2,C)$ and
$\Lambda$ be a homomorphism of $SL(2,C)$ onto $SO(3,1)$. We
consider the one particle states of a particle described by a wave
function in the momentum space transforming under an irreducible
unitary representation of the Poincare group
\cite{wigner}\cite{wein}\cite{novo}\cite{ohnuki}
\begin{equation}
U(A,a)\psi(p)\equiv\psi(A,a;p)=\exp(ipa)D^{j}(A(p))\psi(\Lambda(A^{-1})p),
\end{equation}
where $D^{j}$ is a $(2j+1)$-dimensional irreducible unitary
 representation of $SU(2)$.
The Wigner rotation matrix $A(p)$ in eq.(1) is defined as
\begin{displaymath}
A(p)=\alpha^{-1}(p^{\prime})A\alpha(p),
\end{displaymath}
where $p^{\prime}=\Lambda(A)p$ and the boost $\Lambda(\alpha(p))$
of a particle with mass $m$ is defined by
$\Lambda(\alpha(p))(mc,0,0,0)=p$ ( $c$ denotes the velocity of
light).

The wave function in the position space is defined as usual by the
Fourier transform
\begin{equation}
\psi(x)=\int  d{\bf p}\exp(-ipx)\psi(p),
\end{equation}
where $px=p^{\mu}x_{\mu}=p_{0}x_{0}-{\bf p}{\bf x}$ and  $p_{0}=\sqrt{{\bf
p}^{2}+m^{2}c^{2}}$ (Greek indices run from $0$ t0 $3$, Latin indices from $1$ to $3$,
boldface letters denote three-dimensional vectors). The Fourier transform of $U(A,a)\psi$ is a
non-local function of $x$ because $A(p)$ depends on $p$.

 The representation (1) is unitary in
 the Hilbert space of functions
 square integrable with respect to $d{\bf
p}p_{0}^{-1}$. There are two Casimir operators characterizing
irreducible representations of the Poincare group
$m^{2}=P^{\mu}P_{\mu}$ and the Pauli-Lubanski vector
\begin{displaymath} w_{\mu}w^{\mu}=-m^{2}j(j+1),
\end{displaymath}
where \begin{displaymath} w_{\mu}
=\frac{1}{2}\epsilon_{\mu\nu\sigma\rho}M^{\nu\sigma}P^{\rho}.
\end{displaymath} In the neighborhood of $1$ we write
\begin{displaymath}
A=\exp\Omega,
\end{displaymath}
where $\Omega=i\omega^{\mu\nu}l_{\mu\nu}$, $\omega^{\mu\nu}$ are real parameters
and $l_{\mu\nu}$ is a representation of the Lorentz algebra by $2\times 2$ matrices.
In vector notation
\begin{displaymath} \Omega=i\omega^{rs}l_{rs}+\omega^{0j}k_{j}
=i{\bf a}{\bf l}-{\bf v}{\bf k}.
\end{displaymath}
Here $a_{j}=\epsilon_{jrs}\omega_{rs}$,
$l_{j}=\frac{1}{2}\epsilon_{jrs}l_{rs}$ and $v_{j}=-\omega_{0j}$.
We define the generators $M_{\mu\nu}$ of the unitary
representation of the Lorentz group by
\begin{equation}
U(A)=\exp(i\omega^{\mu\nu}M_{\mu\nu}).
\end{equation}
Let us note that there is some arbitrariness in the definition of
the generators which is a consequence of the non-uniqueness of the
boost. We can choose $\tilde{\alpha}(p)= \alpha(p)g(p)$ where $g
(p)$ is an arbitrary unitary matrix. Then, the Wigner rotation is
\begin{equation}
\tilde{A}(p)=g^{-1}(p^{\prime})A(p)g(p).
\end{equation}
In the calculations of the generators we differentiate eq.(1) over
$\omega^{\mu\nu}$. Then, eq.(4) leads to a gauge transformation of
the generators.

 First, we fix the gauge choosing (this is the
conventional choice \cite{wein}\cite{novo}) $\alpha$ as a
Hermitian matrix. In such a case we obtain from eq.(1)
\begin{equation}
M_{\mu\nu}=L_{\mu\nu}+\Sigma_{\mu\nu}(p),
\end{equation}
where \begin{equation} L_{jk}=-i(p_{j}\frac{\partial}{\partial
p^{k}}-p_{k}\frac{\partial}{\partial p^{j}}),
\end{equation}
\begin{equation} L_{0j}=-ip_{0}\frac{\partial}{\partial
p^{j}}.
\end{equation}
$\Sigma_{\mu\nu}\in su(2)$  results from the differentiation of
$D$ in eq.(1). If we write $J_{l}=\frac{1}{2}\epsilon_{lmn}M_{mn}$
and $K_{l}=M_{l0} $ then after differentiation of eq.(1) we obtain

\begin{equation} J_{r}=-i\epsilon_{rlk}p_{l}\frac{\partial}{\partial
p^{k}}+S_{r},
\end{equation}
\begin{equation} K_{r}=ip_{0}\frac{\partial}{\partial
p^{r}}+\epsilon_{rln}p_{l}S_{n}(p_{0}+mc)^{-1}\equiv
ip_{0}\frac{\partial}{\partial p^{r}}+\Sigma_{0r},
\end{equation}
where
\begin{equation}
[S_{k},S_{l}]=i\epsilon_{kln}S_{n}
\end{equation}
is an irreducible representation of $su(2)$.

Then, the change of the boost (4) leads to a change of the
generators (where $V(g)$ is the $D^{j}$ representation of $g$
appearing in eq.(4))
\begin{equation}
\tilde{\Sigma}_{0r}=V^{-1}\Sigma_{0r}V-V^{-1}L_{0r}V
\end{equation}
whereas $M_{jk}$ does not change because for $A\in SU(2)$ the
Wigner rotation $A(p)$ coincides with $A$.

The wave function (2) transforming under an irreducible
representation of the Poincare group  satisfies the wave equation
\begin{equation}
 (c^{-2}\partial_{t}^{2}-\nabla^{2}+m^{2}c^{2})\psi=0.
\end{equation}

 We consider
random superpositions of states and define the density matrix
\begin{equation} \rho_{\sigma\sigma^{\prime}}(p,p^{\prime})
=\langle\psi_{\sigma}(p)\overline{\psi}_{\sigma^{\prime}}(p^{\prime})\rangle
\end{equation}
and
 \begin{equation} \rho_{\sigma\sigma^{\prime}}(x,x^{\prime})
=\int d{\bf p}d{\bf
p}^{\prime}\exp(-ipx+ip^{\prime}x^{\prime})\langle\psi_{\sigma}(p)\overline{\psi}_{\sigma^{\prime}}(p^{\prime})\rangle .
\end{equation}
Clearly, for a free motion
\begin{equation}
\begin{array}{l}(c^{-2}\partial_{t}^{2}-\nabla^{2}+m^{2}c^{2})\rho(x,x^{\prime})\cr
=(c^{-2}\partial_{t^{\prime}}^{2}-\nabla^{\prime
2}+m^{2}c^{2})\rho(x,x^{\prime})=0.
\end{array}\end{equation}

We define the Wigner function (matrix) $W$ as the Fourier
transform of $\rho$
\begin{equation}\begin{array}{l}
W_{\sigma\sigma^{\prime}}({\bf x},{\bf p}) =\int d{\bf k}d{\bf
k}^{\prime}\int\delta({\bf p}-\frac{1}{2}{\bf k}-\frac{1}{2}{\bf
k}^{\prime}) \exp(i({\bf k}-{\bf k}^{\prime}){\bf
x})\rho_{\sigma\sigma^{\prime}}(k,k^{\prime}).\end{array}
\end{equation}
In the position space
\begin{displaymath}
W_{\sigma\sigma^{\prime}}({\bf x},{\bf p})=\int d{\bf
v}\exp(-i{\bf p}{\bf v}) \rho_{\sigma\sigma^{\prime}}({\bf
x}+\frac{1}{2}{\bf v},{\bf x}-\frac{1}{2}{\bf v}).
\end{displaymath}
Note that from $\rho^{+}=\rho$ it follows that
\begin{equation}
W^{+}=W
\end{equation}and if $\rho\geq 0$ then
\begin{displaymath}
\tilde{W}({\bf p})=\int d{\bf x}W({\bf x},{\bf p})=\rho({\bf
p},{\bf p})\geq 0 .
\end{displaymath}

If an observable $\Phi_{\sigma\sigma^{\prime}}({\bf p}) $ is a
function solely  of ${\bf p}$ then the expectation value in
quantum mechanics can be expressed in the form
\begin{equation}
Tr(\rho \Phi)=Tr\int d{\bf x}d{\bf p}W({\bf p},{\bf x})\Phi({\bf
p}),
\end{equation}
where the trace on the rhs is over the spin indices. The formula
(18) remains valid for observables $\Phi({\bf p},{\bf x})$ on the
phase space. However, the Weyl ordering must be applied  for a
definition of a function of non-commuting operators ${\bf x}$ and
${\bf p}$.

We can also define the density matrix for a calculation of spin
expectation values \begin{displaymath}
\rho_{\sigma\sigma^{\prime}}=\int d{\bf x}d{\bf
p}W_{\sigma\sigma^{\prime}}({\bf x},{\bf p}).\end{displaymath} The
spin matrix in  relativistic quantum mechanics is often discussed
in relation to the EPR experiments \cite{epr}.

The normalization of the probability distribution means
\begin{displaymath}
Tr\int d{\bf x}d{\bf p}W_{\sigma\sigma^{\prime}}({\bf x},{\bf
p})=1 .
\end{displaymath}
 We treat an interaction of a particle with an
environment as a motion through a random medium.
 We assume that the interaction
preserves the Lorentz invariance. In \cite{haba2} we consider the
St\"uckelberg proper time formulation \cite{stuck}\cite{feyn} of
quantum mechanics with a dissipation (see also \cite{hor2}). In
such an approach the free quantum evolution disturbed solely by a
random Lorentz invariant perturbation reads

\begin{equation}\begin{array}{l}
2\kappa^{-2}(\partial_{\tau}\rho+i[(\partial_{0}^{2}-\nabla^{2}+m^{2}c^{2}),\rho])
=M_{\mu\nu}M^{\mu\nu}\rho+\rho M_{\mu\nu}M^{\mu\nu}
-2M_{\mu\nu}\rho M^{\mu\nu} \cr+m^{-2}w_{\mu}w^{\mu}\rho + \rho
m^{-2}w_{\mu}w^{\mu}-2m^{-2}w_{\mu}\rho w^{\mu}
=[M_{\mu\nu},[M^{\mu\nu},\rho]]+m^{-2}[w_{\mu},[w^{\mu},\rho]].\end{array}
\end{equation}
Note that
\begin{displaymath}
[M_{\mu\nu},[M^{\mu\nu},\rho]]
 ={\bf
J}^{2}\rho +\rho{\bf J}^{2}-2{\bf J}\rho{\bf J}  - {\bf K}^{2}\rho
-\rho{\bf K}^{2} +2{\bf K}\rho{\bf K}.
\end{displaymath}
 From eq.(19) it follows that $Tr\rho$ is preserved by the
time evolution  with arbitrary Hermitian operators $M_{\mu\nu}$.
${\bf J}$ and ${\bf K}$ terms enter eq.(19) with different signs.
The $K$ terms are in the Lindblad form \cite{lind}whereas the $J$ terms
are not (there is  a wrong sign in front of $J^{2}$). For this
reason it is not clear whether the dynamics is dissipative. In
sec.3 it will be shown that the dynamics is well-defined and
dissipative if the momentum $p$ is on the mass shell ${\cal H}$

\begin{equation}
p^{2}=p_{0}^{2}-p_{1}^{2}-p_{2}^{2}-p_{3}^{2}=m^{2}c^{2}.
\end{equation}
In the massless case instead of $2j+1$ states of spin $j$ we have
only two states of helicity $\lambda=j$ and $\lambda=-j$.
Eqs.(8)-(9) should be transformed to the helicity basis before the
limit $m\rightarrow 0$ \cite{moses}\cite{foldy}. The formulae for
generators of an irreducible representation of the Poincare group
in the case $m=0$ have been discussed in
\cite{foldy}\cite{shirokov}\cite{birula}. The different form of
the generators results from different realizations of the little
group $E(2)$ of massless particles. We consider the formula of
Bialynicki-Birula \cite{birula} and Shirokov \cite{shirokov}
\begin{equation}
J_{r}=-i\epsilon_{rlk}p_{l}\frac{\partial}{\partial
p^{k}}+\frac{\lambda}{2} \vert{\bf p}\vert\frac{\partial}{\partial
p_{r}}\ln (p_{a}p^{a})
\end{equation}and
\begin{equation} K_{r}=i\vert{\bf p}\vert\frac{\partial}{\partial
p^{r}}-\lambda p_{3}\epsilon_{3rb}p_{b}(p_{a}p^{a})^{-1},
\end{equation} where $a=1,2$ and $\lambda=\vert{\bf p}\vert^{-1}{\bf p}{\bf J}=\vert{\bf p}\vert^{-1}{\bf p}{\bf S}$ is
the helicity equal  $\pm j$ .

For later discussion we mention also the formula of Ohnuki
(\cite{ohnuki} and references quoted there)
\begin{equation}
J_{a}=-i\epsilon_{alk}p_{l}\frac{\partial}{\partial p^{k}}+\lambda
p_{a}( \vert{\bf p}\vert+p_{3})^{-1},
\end{equation}
\begin{equation}
J_{3}=-i\epsilon_{3lk}p_{l}\frac{\partial}{\partial
p^{k}}+\lambda,
\end{equation}
\begin{equation} K_{a}=i\vert{\bf p}\vert\frac{\partial}{\partial
p^{a}}+\lambda \epsilon_{3ab}p_{b}( \vert{\bf
p}\vert+p_{3})^{-1}
\end{equation}and
\begin{equation} K_{3}=i\vert{\bf p}\vert\frac{\partial}{\partial
p^{3}}
\end{equation}

\section{Diffusion equation }
 The equation
for the Wigner function (16) resulting from eq.(19) reads
\begin{equation}\begin{array}{l}
\partial_{\tau}W=p^{\mu}\partial_{\mu}^{x}W+\frac{1}{2}\kappa^{2}(M_{\mu\nu}M^{\mu\nu}W
+WM_{\mu\nu}M^{\mu\nu} -2M_{\mu\nu}WM^{\mu\nu})
\cr
=p^{\mu}\partial_{\mu}^{x}W+\kappa^{2}\Big( {\bf J}^{2}W +W{\bf
J}^{2}-2{\bf J}W{\bf J}
 - {\bf K}^{2}W -W{\bf K}^{2} +2{\bf K}W{\bf K}
 \cr+m^{-2}c^{-2}w_{\mu}w^{\mu}W
 +m^{-2}c^{-2}w_{\mu}w^{\mu}W-2m^{-2}c^{-2}w_{\mu}Ww^{\mu}\Big).
\end{array}
\end{equation}
The derivatives over position will have an index $x$, derivatives
without an index are over momenta.

 The operators ${\bf J}$ and
${\bf K}$ are defined in eqs.(8) and (9). Writing them on the rhs
in eq.(27) means that the differentiation acts as usual on the
function $W$ but the multiplication by the matrix ${\bf S}$ is a
multiplication from the  right. We have for the orbital part
\begin{displaymath}
L_{\mu\nu}W(x,p)=L_{\mu\nu}(x)W+L_{\mu\nu}(p)W.
\end{displaymath}
Unfortunately, the action of $M_{\mu\nu}$ on $W$ is quite
complicated because the spin part depends on the momenta. It is
determined by a non-local integral kernel. We are unable to show
whether the evolution (27) preserves the positivity of $W$ or not.
Eq.(27) simplifies if  we restrict ourselves to the momentum
distribution
\begin{equation} \tilde{W}({\bf p})=\int d{\bf x} W({\bf x},{\bf
p})=\rho({\bf p},{\bf p}).
\end{equation}
In such a case the terms $ L_{\mu\nu}(x) W$ are absent  because $\int d{\bf x} L_{\mu\nu}(x) W =0$. Then,
 eqs.(27) can be rewritten as differential equations (using
eqs.(8)-(9))\begin{equation}\begin{array}{l}
\kappa^{-2}\partial_{\tau}W=\kappa^{-2}p^{\mu}\partial^{x}_{\mu}W+\frac{1}{2}m^{2}c^{2}\triangle_{H}^{m}W-
imc(p_{0}+mc)^{-1}\epsilon_{rlk}p_{l}\frac{\partial}{\partial
p^{k}}[S_{r},W]
 \cr -\frac{1}{2}{\bf
p}^{2}(p_{0}+mc)^{-2}[{\bf S},[{\bf
S},W]]+\frac{1}{2}(p_{0}+mc)^{-2}[{\bf p}{\bf S},[{\bf p}{\bf
S},W]]   \equiv {\cal G}W,\end{array}
\end{equation}
where\begin{equation}
\triangle_{H}^{m}=\partial_{1}^{2}+\partial_{2}^{2}+\partial_{3}^{2}+(mc)^{-2}p_{j}p_{k}\partial^{j}\partial^{k}
+3(mc)^{-2}p_{k}\partial^{k}.
\end{equation}
Eq.(29) is equivalent to eqs.(19) and (27) as an equation for
$\tilde{W}$ (28)(then the term
$p^{\mu}\partial^{x}_{\mu}\tilde{W}=0$). When ${\bf S}=0$ then
eq.(29) coincides with the diffusion equations of Schay \cite{schay} and
Dudley \cite{dudley}\cite{haba}\cite{haba2}. We suggest that eq.(29) defines a proper formulation of the
relativistic  diffusion of particles with spin. Eq.(29) preserves
positivity (as will be shown in the subsequent sections) necessary
for a probabilistic interpretation. In order to preserve the
positivity we rejected in eq.(29) the $L_{\mu\nu}(x)W$ terms from
the dissipation part in eq.(27) (such a term preserves positivity
only if $x$ is restricted to $x^{2}\geq 0$) but we preserved the
term $p^{\mu}\partial_{\mu}^{x}W$ as describing the spatial
Hamiltonian evolution  .

$W$ is a Hermitian matrix. We could decompose it into a basis of
Hermitian matrices. In the case of $j=\frac{1}{2}$
\begin{equation}
W=W^{(0)}+W^{(k)}\sigma_{k},
\end{equation}
where $\sigma^{k}$ are the Pauli matrices and $W^{\mu}$ are real
functions. Eq.(29) can be rewritten as a set of equations
\begin{equation}\begin{array}{l}
\kappa^{-2}\partial_{\tau}W^{(0)}=\kappa^{-2}p^{\mu}\partial^{x}_{\mu}W^{(0)}+\frac{1}{2}m^{2}c^{2}\triangle_{H}^{m}W^{(0)},
\end{array}\end{equation}

\begin{equation}\begin{array}{l}
\kappa^{-2}\partial_{\tau}{\bf
W}=\kappa^{-2}p^{\mu}\partial^{x}_{\mu}{\bf
W}+\frac{1}{2}m^{2}c^{2}\triangle_{H}^{m}{\bf W}\cr
-mc(p_{0}+mc)^{-1}({\bf p}(\partial_{k}W^{(k)})-p_{k}\nabla
W^{(k)}) \cr -\frac{1}{2}{\bf p}^{2}(p_{0}+mc)^{-2}{\bf
W}-\frac{1}{2}(p_{0}+mc)^{-2}({\bf p}{\bf W}){\bf p},\end{array}
\end{equation}
where ${\bf W}=(W^{(1)},W^{(2)},W^{(3)})$ is a spin current
\cite{zhang} defined in general by \begin{displaymath} W^{(k)}=
\frac{1}{2}Tr(\sigma_{k}W).
\end{displaymath}

In the massless case \begin{equation}
\rho_{\lambda\lambda^{\prime}}=\langle
\overline{\psi}(\lambda)\psi(\lambda^{\prime})\rangle .
\end{equation}
Here, $\rho$ and $W$ are $2\times 2$ matrices. The diffusion
equation reads (now $w_{\mu}w^{\mu}=0$, we subtract the helicity
in eq.(27) instead of the spin;we have calculated
$M_{\mu\nu}M^{\mu\nu}$ using either eqs.(21)-(22) or
(23)-(25),both give eq.(35))
\begin{equation}\begin{array}{l}
\kappa^{-2}(\partial_{\tau}W-p^{\mu}\partial^{x}_{\mu}W)=\frac{1}{2}\triangle_{H}W
\end{array}
\end{equation}

with

\begin{equation}
\triangle_{H}=p_{j}p_{k}\partial^{j}\partial^{k}
+3p^{k}\partial_{k},
\end{equation}
where $k=1,2,3$ and $\partial^{j}=\frac{\partial}{\partial
p_{j}}$. We note that the generator (36) \cite{haba2} is the limit
$m\rightarrow 0$ of $m^{2}c^{2}\triangle_{H}$ of ref.\cite{haba}.
The diffusion equation (35) is the limit $m\rightarrow 0$ of  the
diffusion equation (29) if we set ${\bf S}\rightarrow \vert{\bf
p}\vert^{-1}{\bf p}{\bf S}$. Eq.(35) shows that the dissipation
does not depend on the polarization. Its inclusion in
eqs.(34)-(35) is superfluous; the Lorentz invariant dissipation
does not depend on the polarization in the limit $m\rightarrow 0$.
\section{Gauge invariance}
 Let us compare the operator on the rhs of eq.(29) with the
 Laplace-Beltrami operator
 on an $SU(2)$ vector bundle with a connection ${\bf A}$ over the hyperbolic manifold (20)
\begin{equation}
\triangle_{A}=g^{-\frac{1}{2}}(\partial_{j}-iA_{j})g^{\frac{1}{2}}g^{jl}(\partial_{l}-iA_{l}),
\end{equation}
where \cite{haba}
\begin{equation}
g^{jl}=\delta^{jl}+m^{-2}c^{-2}p^{j}p^{l},
\end{equation}
\begin{equation}
g_{jl}=\delta^{jl}-p_{0}^{-2}p^{j}p^{l} \end{equation} and
$g=m^{2}p_{0}^{-2}$ .

We write the diffusion equation as
\begin{equation}
2m^{-2}\kappa^{-2}c^{-2}(\partial_{\tau}W-p^{\mu}\partial^{x}_{\mu}W)=\triangle_{A}W\end{equation}
with
\begin{equation}
{\bf A}=\frac{1}{mc(mc+p_{0})}{\bf p}\times{\bf S}^{ad}
\end{equation}
and \begin{equation}\begin{array}{l} A_{k}g^{kl}A_{l}={\bf
p}^{2}m^{-2}c^{-2}(p_{0}+mc)^{-2}({\bf S}^{ad}{\bf S}^{ad} -{\bf
p}^{-2}{\bf p}{\bf S}^{ad}{\bf p}{\bf
S}^{ad}),\end{array}\end{equation} where
\begin{displaymath} {\bf S}^{ad}\Phi=[{\bf S},\Phi].
\end{displaymath}
A similar connection in the momentum space comes from
Foldy-Wouthuysen transformation in \cite{hall} where it is applied
to a description of the spin Hall effect.

 The diffusion equation
(33) for the spin current ${\bf W}$ takes the gauge invariant form
(40)
\begin{equation}
2m^{-2}\kappa^{-2}c^{-2}(\partial_{\tau}{\bf
W}-p^{\mu}\partial^{x}_{\mu}{\bf W})=\triangle_{A}{\bf
W}.\end{equation} Here, the covariant derivative is $\partial
_{j}+A_{j}$ with an $SO(3)$ (antisymmetric) connection

\begin{equation}
 A^{ab}_{j}=\frac{1}{mc(mc+p_{0})}(p_{b}\delta_{ja}-p_{a}\delta_{jb}),
\end{equation}
where $(A_{j}W)^{a}=A_{j}^{ab}W^{(b)}$ is a multiplication of the
vector ${\bf W}$ by a matrix $A$. Under the gauge transformation
\begin{displaymath}
\tilde{A}_{j}=O^{-1}A_{j}O+O^{-1}\partial_{j}O
\end{displaymath}
and ${\bf W}\rightarrow O^{-1}{\bf W}$. So that eq.(43) stays
invariant.

The diffusion equation (29) defined by $M_{\mu\nu}M^{\mu\nu}$
 depends on the choice of the generators in the Wigner
representation (arbitrariness of the Wigner rotation $A(p)$
resulting from the non-uniqueness of the boost). The arbitrariness
of the generators is expressed by the gauge transformation of the
connection ${\bf A}$
\begin{equation}
\tilde{A}_{j}=V^{-1}A_{j}V+iV^{-1}\partial_{j}V.
\end{equation}
 The gauge (41) is
distinguished by the transversality condition
\begin{displaymath}
g^{jk}\partial_{j}A_{k}=0.
\end{displaymath}
Note that the gauge transformation
 (45) is
different from the one of eq.(11) by a factor of $p_{0}$. We would
get the gauge transformation (11) if we assumed the diffusion
generator in the form $M_{0j}M_{0j}$. The diffusion would have the
gauge invariance (11) but no relativistic invariance. The
subtraction of $M_{jk}M_{jk}$ changes the gauge transformation.
The change comes from the replacement of the
 $p_{0}^{2}\partial_{j}\partial_{j}$  term by
$m^{2}c^{2}\partial_{j}\partial_{j}$ resulting from the
subtraction of $M_{jk}M_{jk}$ from $M_{0j}M_{0j}$. Clearly, this
is also necessary for the Lorentz invariance.

 If in eq.(40) we perform the transformation (45) of the connection
 then after the transformation  the solution of eq.(40) is
rotated  into $V^{-1}W_{\tau}V$. A calculation of expectation
values depends on this gauge rotation.
Only gauge independent expectation values have a physical meaning.
The expectation value is expressed by the trace (18). Hence, as
local observables $\Phi$ we should consider either a multiple of
an identity matrix or a local function of
\begin{equation}
R_{jk}=\partial_{j}A_{k}-\partial_{k}A_{j}+i[A_{j},A_{k}]
\end{equation}
and its covariant derivatives. As non-local observables  we could
admit polynomials of the Wilson loop variables\begin{equation}
\Phi_{C}=\prod_{C}\exp ( iA_{j}dp_{j}),
\end{equation}
where the rhs of eq.(47) denotes a product integral around a closed curve $C$.

\section{Feynman-Kac-Ito formula}

We derive a probabilistic solution of eq.(29). Let us define the
path ordered phase factor (a matrix)as a solution of the equation
\begin{equation} dT_{\tau}^{ad}=iA_{l}\circ
dp^{l}_{\tau}T_{\tau}^{ad} ,\end{equation} where ${\bf p}_{\tau}$
is the stochastic process on the mass-shell (20) discussed in
\cite{haba}(the circle denotes the Stratonovitch integral
\cite{ikeda}). We could  describe the stochastic process as the
diffusion process solving the diffusion equation (32)
\begin{displaymath}
W_{\tau}^{(0)}(x,{\bf p})=E[W^{(0)}(x_{\tau},{\bf p}_{\tau})],
\end{displaymath}
where the expectation value $E[...]$ is over the diffusion process
$ (x_{\tau},{\bf p}_{ \tau})$ as has been discussed in detail in
\cite{haba}.The solution $T_{\tau}$ of eq.(48) is a product
integral (a unitary matrix)\cite{habad}\begin{equation}
T_{\tau}^{ad}=\prod^{s=\tau}_{s=0}\exp(iA^{ad}_{j}\circ dp^{j}_{s}).
\end{equation} Then, the solution of the matrix
equation (29) reads\begin{equation}\begin{array}{l}
W_{\tau}(x,{\bf p})=E[T_{\tau}^{ad}W(x_{\tau},{\bf p}(\tau))] =
E[T_{\tau} W(x_{\tau},{\bf p}_{\tau})T_{\tau}^{-1}],\end{array}
\end{equation} where in the last formula the matrices are in the irreducible representation
(10) of $su(2)$, i.e.,
\begin{displaymath}
{\bf A}=\frac{1}{mc(mc+p_{0})}{\bf p}\times{\bf S}.
\end{displaymath}
The vector equation (43) has the solution
\begin{equation}\begin{array}{l}
{\bf W}_{\tau}(x,{\bf p})=E[T_{\tau}{\bf }{\bf W}(x_{\tau},{\bf
p}_{\tau})] ,\end{array}
\end{equation}
where $T_{\tau}\in O(3)$ is the product integral with the
connection (44). Eqs.(50)-(51) give the solutions of the diffusion
equations (32)-(33).

As a consequence we obtain the diamagnetic inequality
\begin{displaymath}
\vert W_{\tau}(x,{\bf p})\vert \leq E[\vert W(x_{\tau},{\bf
p}(\tau))\vert],
\end{displaymath}
i.e., the probability distribution of particles with a spin is
bounded by the probability distribution of spinless particles
discussed in \cite{haba}. From eq.(50) it follows that if the
initial value $W$ is a Hermitian positive definite matrix then
$W_{\tau}$  is also a Hermitian positive definite matrix, i.e.,
\begin{displaymath}
\int d{\bf p}\overline{f}_{\sigma}({\bf
p})W_{\sigma\sigma^{\prime}}f_{\sigma^{\prime}}({\bf p})\geq 0.
\end{displaymath}
Hence, $W$ can be given a probabilistic interpretation which is
preserved in time.

As discussed in sec.4 the expectation values of ${\bf S}$ are not
gauge invariant. For this reason ${\bf S}$ in not an appropriate
variable for a spin. We should use
\begin{equation}
\hat{S}_{j}=2m^{2}c^{2}\epsilon_{jkl}R_{kl}.
\end{equation}
Then, the expectation value of $\hat{{\bf S}}$ is
\begin{displaymath}\begin{array}{l}
\langle\hat{S}_{j}\rangle_{\tau}=2m^{2}c^{2}\epsilon_{jkl}Tr\int
d{\bf p}E[T_{\tau}^{-1}R_{kl}({\bf p})T_{\tau}W({\bf
p}_{\tau}({\bf p}))].
\end{array}\end{displaymath}
It follows that the expectation value is gauge invariant. It can
also be seen from eq.(50) that (in the absence of an
electromagnetic field) the spin evolution is just a spin rotation
determined by the connection ${\bf A}$ (41). By differentiation of
$T_{\tau}$ in $\langle\hat{S}_{j}\rangle_{\tau}$ using (48) and
the Ito formula \cite{ikeda}, we obtain (see also eq.(29)) the
double commutator
\begin{displaymath}
\partial_{\tau}\langle\hat{S}_{j}\rangle_{\tau}\simeq [{\bf
S},[{\bf S},\langle\hat{S}_{j}\rangle_{\tau}]],
\end{displaymath}
where ${\bf S}$ is the matrix representation of $su(2)$. Such a
dissipative spin dynamics is discussed in \cite{muller}.

\section{Interaction with an electromagnetic field }  Without
external fields the spin is completely described by the
Pauli-Lubanski vector
 (in the particle's rest frame $w=(0,m{\bf s})$).
 It does not change in time.In an  electromagnetic field the
amplitudes do not satisfy the Klein-Gordon equation (12).The
standard way to include the interaction with an external
electromagnetic field is to define bispinor amplitudes which
satisfy the Dirac equation (the space-time derivatives replaced by
covariant derivatives) .  The bispinor amplitudes describe both a
particle and the antiparticle.
  In an external field the division
 of Dirac wave functions into positive and negative frequencies
 (electrons and positrons) is field-dependent and non-covariant.
We need to eliminate the negative frequencies. We do it in an
expansion in $c^{-1}$. Then, the gauge invariance in momentum
space will be our guiding principle for a derivation of the final
relativistic diffusion equation. We express the lower bispinor by
the upper one. Then,  we expand the field-dependent parts of the
Dirac equation in powers of $c^{-1}$. In the lowest non-trivial
order the square of the Dirac operator for a particle in an
electromagnetic vector potential ${\bf a}$ ( determining the
magnetic field ${\bf B}=\nabla\times{\bf a}$) and electric field
${\bf E}$ can be expressed in the form \cite{anadan}\cite{jurg}
\begin{equation}
D^{2}=\partial_{0}^{2}-\nabla_{k}\nabla_{k}+\frac{e}{mc}{\bf
S}{\bf B},\end{equation} where \begin{displaymath}
\nabla_{k}=\partial_{k}-ia_{k}+\frac{i}{4mc^{2}}\epsilon_{kjl}E_{j}S_{l}.
\end{displaymath}
As discussed in \cite{anadan}\cite{jurg} we obtain an extended $U(1)\times SU(2) $ gauge invariance
in the position space as a result of the coupling to the spin
degrees of freedom. The Pauli equation is expressed as the
Klein-Gordon equation in a non-Abelian gauge field with the
symmetry group $U(1)\times SU(2)$.

In the proper time approach to the density matrix evolution we
insert the Pauli approximation (53) of the square $D^{2}$ of the
Dirac operator in the commutator with the density matrix in
eq.(19). Then,  the Wigner function evolution is determined by the
motion of position and momenta determined by the equations

 \begin{displaymath} \frac{dx^{\mu}}{d\tau}=\frac{p^{\mu}}{m},
\end{displaymath}
\begin{equation}
\frac{dp_{j}}{d\tau}=\frac{e}{mc}F_{j\mu}p^{\mu}.
\end{equation} The equation for the spin
comes from the commutator $[D^{2},\rho]$ in the master equation
(19). In the classical non-relativistic limit this equation reads
(see a discussion in
\cite{bmt}\cite{keller}\cite{fradkin}\cite{spohn} and a
generalization to non-Abelian gauge fields in
\cite{arodz}\cite{heinz}\cite{kelly})

\begin{equation}
\frac{dS_{j}}{d\tau}=\frac{e}{mc}\epsilon_{jkl}(B_{k}-\frac{1}{2mc}\epsilon_{krn}p_{r}E_{n})S_{l}.
\end{equation}  In the approximation (53)-(55) to eq.(19) the equation (29) for the
evolution of the Wigner function  in an electromagnetic fields
takes the form
\begin{equation}\begin{array}{l}
\kappa^{-2}\partial_{\tau}W=\kappa^{-2}p^{\mu}\partial^{x}_{\mu}W+\frac{1}{2}m^{2}c^{2}\triangle_{H}^{m}W-
imc(p_{0}+mc)^{-1}({\bf p}\times\nabla)[{\bf S},W]
 \cr -\frac{1}{2}{\bf
p}^{2}(p_{0}+mc)^{-2}[{\bf S},[{\bf
S},W]]+\frac{1}{2}(p_{0}+mc)^{-2}[{\bf p}{\bf S},[{\bf p}{\bf
S},W]] \cr +\kappa^{-2}\frac{ie}{mc}({\bf B}-\frac{1}{2mc}{\bf
p}\times {\bf E})[{\bf S},W]+ \kappa^{-2}\frac{e}{m}({\bf
E}\nabla+\frac{1}{mc}{\bf B}({\bf p}\times \nabla))W.\end{array}
\end{equation}
We need to add terms of higher orders in $c^{-1}$ in order to
obtain an equation which will be gauge invariant in the momentum
space and have the correct Thomas precession
\cite{bmt}\cite{fradkin}\cite{spohn} of the spin (it is surprising
that the $U(1)\times SU(2)$ gauge symmetry in the position space
contributes to the gauge invariance in the momentum space, see
also \cite{hall}).

We suggest that the  gauge invariant version of eq.(56) with the
correct classical Thomas precession is
\begin{equation}\begin{array}{l}
\partial_{\tau}W-p^{\mu}\partial^{x}_{\mu}W=\frac{\kappa^{2}}{2}m^{2}c^{2}\triangle_{A}W\cr
+\frac{ie}{2m^{3}c^{3}}({\bf B}-\frac{p_{0}}{mc(mc+p_{0})}{\bf
p}\times {\bf E})[\hat{{\bf
S}},W]+\frac{e}{mc}F_{j\mu}p^{\mu}(\partial_{j}-iA_{j})W.
\end{array}
\end{equation}
In eq.(57) the gauge covariant spin $\hat{{\bf S}}$ (52) has been
introduced. In the lowest orders in $c^{-1}$ only its
non-relativistic version ${\bf S}$ (as in eq.(56)) appears. The
term $p_{0}(p_{0}+mc)^{-1}$ comes  from the relativistic theory of
spin precession  \cite{bmt}-\cite{spohn} (replacing $\frac{1}{2}$
of eq.(56)).
\section{Spin diffusion}

In this section we show that eqs.(29)-(33) can be equivalently
treated as equations for a function $W(x,p,{\bf n})$ defined on an
extended phase space $R^{4}\times {\cal H}\times S^{2}$ (${\bf
n}\in S^{2}$). Let $g\in SU(2)$ be represented in the form
\begin{equation}
g(\phi,\theta,\alpha)=g_{3}(\phi)g_{2}(\theta)g_{3}(\alpha),
\end{equation}
where $g_{3}$ corresponds (under the $SU(2)\rightarrow O(3)$
homomorphism) to the rotation with respect to the third axis and
$g_{2}$ corresponds to the rotation around the second axis. We can
represent a point ${\bf n}\in S^{2}$ as ${\bf
n}(\phi,\theta)=O(g(\phi,\theta,\alpha)){\bf n_{0}}$ where ${\bf
n}_{0}$ is parallel to the third axis and $O\in SO(3)$. Define
\begin{equation}
Y_{j\sigma}(\theta,\phi)=D^{j}_{\sigma 0}(g(\phi,\theta,\alpha)).
\end{equation}
For natural $j$ these are the standard spherical functions. Let
$W$ be the  Wigner $(2j+1)\times (2j+1)$ matrix  (16). We define a
real function on the extended phase space
\begin{equation}
W(x,{\bf p},{\bf
n})=\sum_{\sigma\sigma^{\prime}}\overline{Y}_{j\sigma}(\phi,\theta)W_{\sigma\sigma^{\prime}}(x,{\bf
p})Y_{j\sigma^{\prime}}(\phi,\theta).
\end{equation}
In the simplest case $j=\frac{1}{2}$ it can be checked by
elementary calculations that if $W$ is of the form (31) then the
transformation (60) gives\begin{equation} W=W^{(0)}+n_{k}W^{(k)},
\end{equation}
where $n_{3}=\cos\theta,n_{1}=\sin\theta\cos\phi,
n_{2}=\sin\theta\sin\phi$. Hence, the expansion (61) in $n_{k}$ is
equivalent to the expansion (31) in the basis of Pauli matrices
$\sigma_{k}$. In the representation (60) the spin operator is
represented as a differential operator
\begin{equation}
S_{k}=-i\epsilon_{klr}n^{l}\frac{\partial}{\partial n^{r}},
\end{equation} where ${\bf n}^{2}=1$. The
operators ${\bf S}$ of eq.(62) are the generators of rotations on
the unit sphere. They could be represented in the spherical
coordinates $(\theta,\phi)$ in the form well-known from the theory
of angular momentum in quantum mechanics.

 The gauge transformation
(11) is just a rotation of ${\bf n}$,i.e., with another
realization of the Wigner rotation, when the solution of the
matrix diffusion equation (29)transforms as
\begin{equation}
W^{\prime}=V(g)^{-1}WV(g),
\end{equation}
we have
\begin{equation}
W^{\prime}({\bf p},{\bf n})=W({\bf p},O(g){\bf n}),
\end{equation}
where $O(g)$ is the rotation corresponding to the element $g\in
SU(2)$ (it may depend on ${\bf p}$).

 We can treat eq.(29) as a diffusion equation on the extended
phase space.  The spin evolution (55)  can equivalently be
described as an evolution of a point on the unit sphere
\begin{equation}
\frac{d{\bf n}}{d\tau}=\frac{e}{mc}({\bf B}\times{\bf
n}-\frac{1}{2mc}({\bf p}\times{\bf E})\times{\bf n}).\end{equation}
It leads to the drift
\begin{displaymath}
Y^{S}W =\frac{e}{mc}\Big({\bf B}-\frac{1}{2mc}{\bf p}\times{\bf
E}\Big)({\bf n}\times\nabla_{{\bf n}})W\end{displaymath} in the
diffusion equation. The spin diffusion part
\begin{equation}\begin{array}{l}
\triangle_{S}={\bf p}^{2}m^{-2}c^{-2}(p_{0}+mc)^{-2}({\bf S}^{ad}{\bf
S}^{ad} -{\bf p}^{-2}{\bf p}{\bf S}^{ad}{\bf p}{\bf
S}^{ad})\end{array}\end{equation} is expressed as a differential
operator
\begin{equation}\begin{array}{l}
\triangle_{S}={\bf p}^{2}m^{-2}c^{-2}(p_{0}+mc)^{-2}\Big(({\bf
n}\times\nabla_{{\bf n}})^{2} -{\bf p}^{-2}({\bf p}({\bf
n}\times\nabla_{{\bf n}}))^{2}\Big).\end{array}\end{equation} The
term
\begin{equation}\begin{array}{l}-
imc(p_{0}+mc)^{-1}\epsilon_{rlk}p_{l}\frac{\partial}{\partial
p^{k}}[S_{r},W]\cr =mc(p_{0}+mc)^{-1} ({\bf n}\times\nabla_{{\bf
n}})({\bf p}\times\nabla)W \end{array}\end{equation} describes the
spin-orbit interaction. Eqs.(29) and (66)-(68) altogether define a
diffusion on the product of the particle phase space with a unit
sphere.

 The
complete diffusion equation of the particle with spin in magnetic
and electric fields equivalent to eq.(56) (with a non-relativistic
spin precession) reads

\begin{equation}\begin{array}{l}
\kappa^{-2}\partial_{\tau}W=\kappa^{-2}p^{\mu}\partial^{x}_{\mu}W+\frac{1}{2}m^{2}c^{2}\triangle_{H}^{m}W
 +\kappa^{-2}\frac{e}{mc}({\bf B}
-\frac{1}{2mc}{\bf p}\times{\bf E})({\bf n}\times\nabla_{{\bf
n}})W\cr+\kappa^{-2}\frac{e}{mc}({\bf p}\times {\bf B}+mc{\bf
E})\nabla W+mc(p_{0}+mc)^{-1} ({\bf n}\times\nabla_{{\bf
n}})({\bf p}\times\nabla)W\cr+\frac{{\bf
p}^{2}}{2m^{2}c^{2}(p_{0}+mc)^{2}}\Big(({\bf n}\times\nabla_{{\bf
n}})^{2} -{\bf p}^{-2}({\bf p}({\bf n}\times\nabla_{{\bf
n}}))^{2}\Big)W.
\end{array}\end{equation}
 In sec.9 we still add a friction drift term
to eq.(69)which drives the diffusion to an equilibrium. A solution
$W_{\tau}(x,{\bf p},{\bf n})$ of the diffusion equation (69)
starting from a non-negative initial condition $W$ remains
non-negative. We can relate the function $W$ to the matrix
$W_{\sigma\sigma^{\prime}}$ (defined as a Fourier transform of the
density matrix (16)) using the expansions (31) and (60). On the
extended phase space the expectation values are defined as in
classical mechanics
\begin{equation}
\langle \Phi({\bf p},{\bf n})\rangle_{\tau}=\int d{\bf p}d{\bf
n}W_{\tau}({\bf p},{\bf n})\Phi({\bf p},{\bf n}),
\end{equation}
where $d{\bf n}$ is the invariant measure on the sphere. The gauge
invariance means the rotation invariance of the integral
(70),i.e., it should not depend on $O({\bf p})$ if we make a
replacement ${\bf n}\rightarrow O({\bf p}){\bf n}$.

Eq.(69) is non-trivial even in the non-relativistic
limit\begin{equation}\begin{array}{l}
\kappa^{-2}\partial_{t}W=\kappa^{-2}{\bf p}\nabla_{{\bf
x}}W+\frac{1}{2}m^{2}c^{2}\triangle W\cr +\frac{1}{8} {\bf
p}^{2}m^{-4}c^{-4}\Big(({\bf n}\times\nabla_{{\bf n}})^{2} -{\bf
p}^{-2}({\bf p}({\bf n}\times\nabla_{{\bf n}}))^{2}\Big)W  +
\frac{1}{2}({\bf n}\times\nabla_{{\bf n}})({\bf
p}\times\nabla)\Big)W\cr +\frac{e}{mc\kappa^{2}}{\bf B}({\bf
n}\times\nabla_{{\bf n}})W-\frac{e}{2m^{2}c^{2}\kappa^{2}}({\bf
p}\times{\bf E})({\bf n}\times\nabla_{{\bf n}})W
+\frac{e}{mc\kappa^{2}}{\bf p}\times {\bf B}\nabla
W+\frac{e}{m\kappa^{2}}{\bf E}\nabla W,\end{array}\end{equation}
where
$\triangle=\partial_{1}^{2}+\partial_{2}^{2}+\partial_{3}^{2}$.

\section{Evolution of observables} The mean values in
quantum mechanics are defined in eq.(18) with $W$ as the Wigner
function. We extend this definition to the diffusion
\begin{equation}\begin{array}{l}
\langle \Phi\rangle_{\tau} =Tr\int dxd{\bf p} W_{\tau}({\bf
p},x)\Phi({\bf p},x) \equiv Tr\int dx d{\bf p} W({\bf
p},x)\Phi_{\tau}({\bf p},x).\end{array}\end{equation} If the
observable $\Phi$ does not depend on $x$ then the formula (72)
coincides with the one of quantum mechanics defining the
expectation values by the trace over the density matrix (the lhs
of eq.(18)). Eq.(72) defines expectation values in the theory of
diffusion processes. $\Phi_{\tau}$ on the rhs of eq.(72) is the
definition of the time evolution of $\Phi$ (the adjoint of the operator
$W\rightarrow W_{\tau}$). We suggest that
quantum expectation values are approximated by the expectation
values (72) taken over the relativistic diffusion process.

As discussed in sec.2 the form of the diffusion equation depends
on the choice of the Wigner rotation. The expectation values (72)
will not depend on the choice of the Wigner rotation if the observable $\Phi$ is
covariant under the gauge transformation (45). This means that
$\Phi$ should either be a scalar or transform as $V^{-1}\Phi V$
under the gauge transformation. In the latter case $\Phi$ can be
 a function of  the curvature $R$ (and its covariant derivatives)
(46) or the loop $\Phi_{C}$ (47). Hence, we must work with
variables $(x,{\bf p},\hat{{\bf S}})$ (52) instead of $(x,{\bf
p},{\bf S})$. The difference appears only in the relativistic
domain because if $\vert {\bf p}\vert <<mc$ then
\begin{displaymath}
R_{jk}\simeq
(2m^{2}c^{2})^{-1}\epsilon_{jkl}S_{l}=(2m^{2}c^{2})^{-1}\Sigma_{jk}.
\end{displaymath}
If we choose the boost (4) with the rotation $g$ then the solution
of the modified diffusion equation will be $V(g)^{-1}W_{\tau}V(g)$
(where $W_{\tau}$ is the solution of the diffusion equation (29)
with the Hermitian boost). As an example, according to
eqs.(45)-(46) the gauge transformed observable $\Phi(\tilde{\hat{{\bf
S}}})$ being a local function of  $\hat{{\bf S}}$ is
transformed into $V^{-1}\Phi(\hat{{\bf S}})V$. Hence, the trace in
eq.(72)  does not depend on the choice of the Wigner rotation.

From eq.(50) it follows that the diffusion with spin is just a
unitary  rotation of the spin followed by the spinless evolution
on the phase space, i.e.,
\begin{equation}\begin{array}{l}
\langle \Phi_{\tau}\rangle=\int d{\bf p}dx E[T_{\tau}^{ad}W({\bf
p}_{\tau},x_{\tau})\Phi({\bf p},x)] \cr = \int dxd{\bf
p}E[T_{\tau}^{-1} \Phi({\bf p},x)T_{\tau}W({\bf p}_{\tau},
x_{\tau}) ],\end{array}
\end{equation}
where $({\bf p}_{\tau},x_{\tau})$ is the solution of the same
diffusion equations as discussed earlier \cite{haba} in the model
without the spin.

The evolution $\Phi_{\tau}$ of $\Phi$ in eq.(72) is determined by
the adjoint operator ${\cal G}^{*}$
\begin{equation}\begin{array}{l}
\kappa^{-2}\partial_{\tau}\Phi\equiv {\cal
G}^{*}\Phi=-p^{\mu}\partial^{x}_{\mu}\Phi+\frac{1}{2}m^{2}\triangle_{H}^{m*}\Phi+im\frac{\partial}{\partial
p^{k}} (p_{0}+mc)^{-1}\epsilon_{rlk}p_{l}[S_{r},\Phi] \cr
-\frac{1}{2}{\bf p}^{2}(p_{0}+mc)^{-2}[{\bf S},[{\bf
S},\Phi]]+\frac{1}{2}(p_{0}+mc)^{-2}[{\bf p}{\bf S},[{\bf p}{\bf
S},\Phi]]  , \end{array}
\end{equation}
where
\begin{equation}
\triangle_{H}^{m*}=\partial_{1}^{2}+\partial_{2}^{2}+\partial
_{3}^{2}+(mc)^{-2}\partial^{j}\partial^{k}p_{j}p_{k}
-3(mc)^{-2}\partial_{k}p^{k}.
\end{equation}
We have two candidates for time in the relativistic diffusion:
$\tau$ (interpreted as the proper time ) and $x_{0}$ . In the
non-relativistic limit , when $p_{0}\simeq mc$, $x_{0}$ and $\tau$
enter the solutions of diffusion equations in an additive way.
Hence, $x_{0}$ is just a shift of $\tau$. In the relativistic case
in order to obtain a solution of the diffusion equation as a
function of the laboratory time $x_{0}$ we let $\tau\rightarrow
\infty$ (see the discussion in \cite{haba}-\cite{haba2}). In the
limit we obtain solutions of eq.(74) which are independent of
$\tau$ . The solution of eq.(29)( or more general eq.(57),
expressed as $\partial_{\tau}W={\cal G}W$ ) does not  depend on
$\tau$ if
\begin{equation}
{\cal G}W=0.
\end{equation}
$\Phi$ does not depend on $\tau$ if
\begin{equation} {\cal G}^{*}\Phi=0.
\end{equation}
Eq.(76) is a transport equation in the laboratory time $x_{0}$
which is well-defined for $x_{0}\leq 0$ ( because it is of the
form $\partial_{0}W=-\triangle W$ where $-\triangle$ is a
positively definite operator). Eq.(77) is a well-defined transport
equation for $x_{0}\geq 0$
 ( because it is of the
form $\partial_{0}\phi=\triangle\phi$). If either of eqs.(76)-(77)
is satisfied then the expectation value (72) does not depend on
$\tau$.
 We treat eq.(77) as the basic formula determining
 the evolution  of observables  measured in
 experiments. Solutions of eqs.(77) and (57) can be related
  by a random change of time which is a generalization of the
 transformation between the evolution in proper time and
 laboratory time $x_{0}$ well-known from the relativistic
 classical mechanics. In  \cite{haba2} we have shown that from
 solutions of  eq.(57)
 (without spin) by a random change of time \cite{ikeda} we obtain solutions of
 the transport eq.(77). In ref. \cite{dun} the problem is discussed in a mathematically equivalent form
 but in the inverse direction:
 from the solutions of the transport equation (77) by a random change of time the authors derive the solution of the evolution equation in the proper time.

\section{An approach to the equilibrium }
We are interested whether the solutions of eqs.(57),(74),(76) or
(77) tend to a limit (the equilibrium) at $\tau\rightarrow \infty$
and $x_{0}\rightarrow \pm\infty$.
In order to achieve the equilibrium additional drift terms ( a
friction) must be added to the diffusion equation. We add a drift
\begin{equation}
Y=R_{j}\frac{\partial}{\partial p_{j}}
\end{equation}
to the diffusion (29). Then
\begin{equation}
{\cal G}^{Y}={\cal
G}+Y=\frac{1}{2}\kappa^{2}m^{2}c^{2}\triangle_{A}+Y.
\end{equation}
 Now, the diffusion equation reads\begin{equation}
\partial_{\tau}W_{\tau}={\cal G}^{Y}W_{\tau}.\end{equation}

The  equation for $\Phi$ is
\begin{equation}
\partial_{\tau}\Phi_{\tau}={\cal G}^{Y*}\Phi_{\tau}.
\end{equation}
 The transport equation (77) reads
\begin{equation}\begin{array}{l}
\kappa^{-2}p_{0}\partial_{0}\Phi=\kappa^{-2}{\bf p}\nabla_{{\bf
x}}\Phi+\frac{1}{2}m^{2}c^{2}\triangle_{H}^{m*}\Phi\cr-imc\frac{\partial}{\partial
p^{k}} (p_{0}  +mc)^{-1}\epsilon_{rlk}p_{l}[S_{r},\Phi]\cr
-\frac{1}{2}{\bf p}^{2}(p_{0}+mc)^{-2}[{\bf S},[{\bf
S},\Phi]]+\frac{1}{2}(p_{0}+mc)^{-2}[{\bf p}{\bf S},[{\bf p}{\bf
S},\Phi]] -\partial_{j}R^{j}\Phi . \end{array}
\end{equation}
Eq.(82) could also be written as
\begin{equation}\begin{array}{l}
\kappa^{-2}p_{0}\partial_{0}\Phi=\kappa^{-2}{\bf p}\nabla_{{\bf
x}}\Phi+\frac{1}{2}m^{2}c^{2}\triangle(-A)\Phi
-\partial_{j}R^{j}\Phi,
\end{array}
\end{equation}
where $\triangle(-A)$ means $\triangle_{A}$ with ${\bf A}\rightarrow -{\bf A}$.

 If $R^{j}$ is a function (i.e. a multiple of the unit matrix) and
we assume that $\Phi$ is a multiple of the unit matrix
 then the transport and equilibrium equations for $\Phi$
 are the same as in the spinless case \cite{haba}.
It follows that
we obtain the  J\"uttner equilibrium distribution \cite{jut}
\begin{equation}
 \Phi_{EJ}=\exp(-\beta p_{0}c)
 \end{equation}
 if
 \begin{equation}
 Y=(-\frac{1}{2}\kappa^{2}\beta cp_{0}p^{j}+\frac{1}{2}\kappa^{2}p^{j})\frac{\partial}{\partial
 p^{j}}.
 \end{equation}
For modified J\"uttner distribution \cite{dun}\cite{cub}
\begin{equation}
 \Phi^{M}_{EJ}=p_{0}^{-1}\exp(-\beta p_{0}c)
 \end{equation}
we have
\begin{equation}
 Y^{M}=-\frac{1}{2}\kappa^{2}\beta cp_{0}p^{j}\frac{\partial}{\partial
 p^{j}}.
 \end{equation}

The equilibrium distribution (86) does not depend on the spin
variables. We know magnetized   systems which have a non-uniform
spin (magnetic moment) distribution. We can obtain such a
distribution if ${\bf B}=const \neq 0$ and ${\bf E}=0$. First, let
us note that if there is no spin diffusion ( $\kappa=0$ ) then
\begin{equation}
W_{B} ({\bf p},{\bf n})=p_{0}^{-1}\exp(-\beta (p_{0}c+{\bf B}{\bf
n}))
\end{equation}
is the solution of eq.(76) with the generator ${\cal G}$ defined by
the rhs of eq.(69). The spin diffusion ($\kappa\neq 0$ )also has
(88) as the equilibrium distribution if instead of the drift (87) we
choose
\begin{equation}\begin{array}{l}
 Y=-\frac{1}{2}\kappa^{2}\beta cp_{0}{\bf p}\nabla
 +mc\beta\kappa^{2}(p_{0}+mc)^{-1}
({\bf n}\times{\bf B})({\bf p}\times\nabla)\cr
 +\beta\kappa^{2}\frac{1}{2} {\bf
p}^{2}m^{-2}c^{-2}(p_{0}+mc)^{-2}\Big(({\bf n}\times{\bf B})({\bf n}\times\nabla_{{\bf n}}) \cr
-{\bf
p}^{-2}{\bf p}({\bf n}\times {\bf B}){\bf p}({\bf n}\times\nabla_{{\bf n}})\Big).\end{array} \end{equation}

The model (88)-(89) may have applications to a description of
the time evolution of magnetic properties of relativistic systems
with spin.

\section{Discussion}

We have discussed an extension of the relativistic diffusion of
Schay \cite{schay} and Dudley \cite{dudley} to particles with
spin. The extension is uniquely determined by the rule that
similarly as in the spinless case the diffusion should be
generated by $M_{\mu\nu}M^{\mu\nu}$. We obtain a diffusion which
is the unique relativistic diffusion involving both the momenta
and the spin. The diffusion defines the Lindblad dissipative
dynamics of particles with spin. This can be an interesting model
for a study of the role of spin in the particle motion in the
phase space. The relevance of the spin-orbit coupling for
dissipation and decoherence has been studied before for  the
non-relativistic velocities  \cite{spintronics}\cite{spinorbit}.
The spin-orbit coupling results from the relativistic theory. The diffusive
dissipation may be relevant to the relativistic description of
entaglement and decoherence in the EPR-type experiments
\cite{epr}. The spin diffusion has been studied before in the
theory of magnetism related to the Landau-Lifshitz equation
\cite{spindiffusion}. However, our relativistic approach
determines the coupling between spin and momenta in the unique
way. It can be useful in a description of dynamic phenomena in the
theory of magnetism. The main field of applications should be in
the realm of relativistic physics. The dissipation resulting from
the synchrotron radiation can be interpreted as a diffusion in a
radiation field . The spin diffusion could be compared with
experimental results of spin rotation in the synchrotron
\cite{synchro}. The relativistic diffusion of particles without
spin has already been applied to heavy ion collisions \cite{ion}.
For more precise experiments when the results are not averaged
over polarizations the description of spin may be relevant.
Finally, in astrophysics the problem of propagation of neutrinos
scattered in the interstellar medium \cite{bernstein}\cite{netrino}
could be considered as a natural domain of application of the
diffusion equation. As follows from eq.(35) for the diffusion of
light (treated as spinless in \cite{haba2}) the spin of the photon
does not play any role.

\end{document}